\documentstyle[twoside,fleqn,espcrc2,epsf]{article}

\title{Quenched QCD with domain wall fermions}
 
\author{
  P.~Chen, N.~Christ, G.~Fleming, A.~Kaehler, C.~Malureanu,
  R.~Mawhinney\thanks{
    R.~Mawhinney presented talk at Lattice '98.  Work supported
    in part by the US Department of Energy.  Calculations were
    done on the QCDSP computer at Columbia University.
  },
  G.~Siegert\thanks{Supported by the Max Kade Foundation},
  C.~Sui, P.~Vranas\thanks{
    Current address:  Physics Department, University of Illinois,
    Urbana, IL 61801
  },
  and Y.~Zhestkov\address{Department of Physics, Columbia University,
  New York, NY 10027, USA }
}
 
\begin{document}

\def\thepage{CU--TP--915}
\thispagestyle{myheadings}

\begin{abstract}

  We report on simulations of quenched QCD using domain wall fermions,
  where we focus on basic questions about the formalism and its ability
  to produce expected low energy hadronic physics for light quarks.
  The work reported here is on quenched $8^3 \times 32$ lattices at
  $\beta = 5.7$ and $5.85$, using values for the length of the fifth
  dimension between 10 and 48.  We report results for parameter choices
  which lead to the desired number of flavors, a study of undamped
  modes in the extra dimension and hadron masses.

\end{abstract}

\maketitle

\section{INTRODUCTION}

Domain wall fermions \cite{kaplan} were originally proposed as a
technique for putting chiral fermions on the lattice.  For fermions
coupled to a vector gauge theory (like QCD), this approach has given
a way to treat massless (or at least controllably small mass) fermions,
while preserving the full continuum symmetry group.  Subsequent
improvements to the original formulation \cite{furman-shamir} made
simulations more practical and studies of the Schwinger model
\cite{vranas}, including dynamical fermion effects, and quenched QCD
\cite{blum-soni} have been very encouraging.  Here we report
a reasonably systematic study of low energy, non-anomalous QCD
physics using domain wall fermions.  For a review and further
references, see \cite{blum}.

\section{BASIC QUESTIONS}

For quenched domain wall simulations, there are three input parameters
for the fermions:  $m_f$, the explicit four-dimensional bare quark
mass;  $L_s$, the extent of the lattice in the fifth dimension and
$m_0$, the five-dimensional bare quark mass.  Except for different
symbols for the three parameters listed above, our conventions follow
\cite{furman-shamir}.

A first basic question involves choosing parameters so that the desired
number of flavors of light fermions is being studied.  For free
fermions, one easily sees that for $m_0 < 0$ there are no light surface
states at the ends of the fifth dimension, for $ 0 < m_0 < 2 $ surface
states for a single light quark flavor appear, for $ 2 < m_0 < 4 $
surface states for four light flavors appear and for $ 4 < m_0 < 6 $
there are surface states for six light quark flavors.  There is a
symmetry under $m_0 \rightarrow 10 - m_0$, which remains when coupling
to gauge fields is introduced, although then these values of
$m_0$ shift.

\begin{figure}[htb]
\epsfxsize=\hsize
\epsfbox{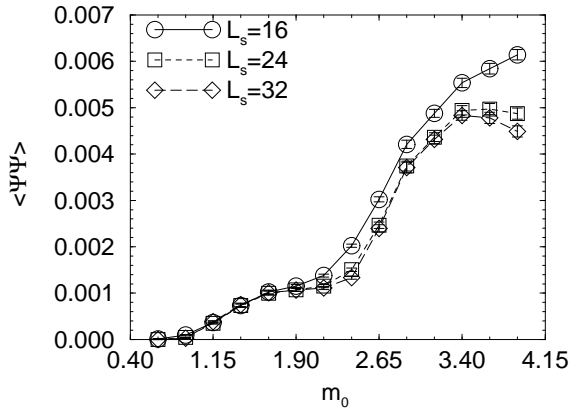}
\caption{The four-dimensional chiral condensate extrapolated to
zero quark mass as a function of the five-dimensional domain
wall mass.}
\label{fig:pbp_vs_m0}
\end{figure}

Figure \ref{fig:pbp_vs_m0} shows one way to probe the number of light
fermions.  The graph is for quenched simulations at $\beta=5.85$ on an
$8^3 \times 32$ lattice.  For a given value of $m_0$, we have measured
the chiral condensate for $m_f = 0.02, 0.04, 0.06, 0.08$ and $0.10$,
and calculated $\langle \bar{\psi} \psi( m_f \rightarrow 0 ) \rangle$
(simple linear fits work well).  This extrapolation was done for three
different values of $L_s$, to make sure we were close to the large
$L_s$ limit.  The figure shows this extrapolated chiral condensate as a
function of $m_0$.

For $m_0 < 0.9 $ the extrapolated chiral condensate is zero, indicating
no spontaneous chiral symmetry breaking, due to the absence of surface
states.  For larger values of $m_0$ the chiral condensate is non-zero
and increasing, until leveling off for $ 1.65 < m_0 < 2.15$.  This
gives a region where there appears to be a single surface state. (For
free fermions, the normalization of the fields depends on $m_0$;  this
region of insensitivity to $m_0$ does not appear to have been
anticipated.)  For larger $m_0$ the chiral condensate increases as the
change from a one flavor to a four flavor theory occurs.  Above about
3.4, a region of relative insensitivity to $m_0$ occurs, where there
are four flavors of light fermions.  The chiral condensate in this
region is numerically very close to four times its value for $1.65 <
m_0 < 2.15$.

\pagenumbering{arabic}
\addtocounter{page}{1}

A second basic question is whether the mixing between the chiral modes
at $s = 0$ and $s= L_s - 1$ can be controlled by choosing $L_s$ large.
It is known that a gauge field with a zero eigenvalue for the
four-dimensional Wilson Dirac operator (with mass $m_0$), has no
damping for modes propagating in the fifth dimension.  We wanted to
explicitly measure the effect of this undamped mode on non-anomalous low
energy QCD physics.  To this end, we have studied its effect on the
chiral condensate (for related work see \cite{fsu}).

The four-dimensional chiral condensate is defined through the
inverse propagator for domain wall fermions, where $s = 0 (L_s-1)$
is a source for right (left) handed quarks and $s=L_s-1 (0)$ is a sink
for left (right) handed quarks.  We have studied a generalized
condensate, where the distance from source to sink is variable.
When this distance is $L_s - 1$, we have the conventional
condensate.

\begin{figure}[htb]
\epsfxsize=\hsize
\epsfbox{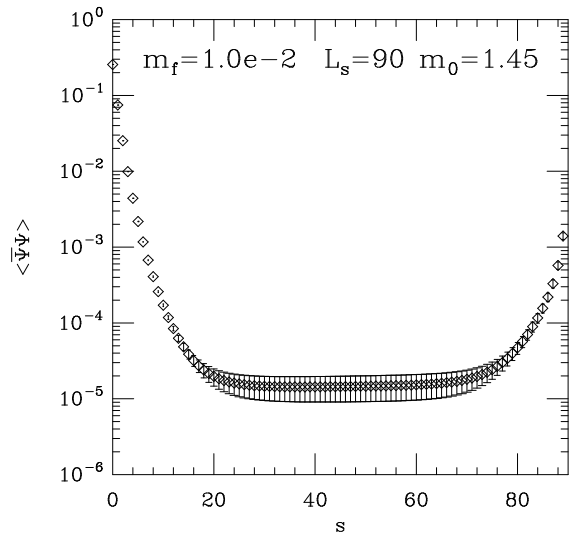}
\caption{A generalized version of the four-dimensional chiral
condensate, where the separation in the fifth dimension between
the source and sink is $s$.  $s = L_s - 1 $ is the usual condensate.}
\label{fig:pbp_vs_ls}
\end{figure}

For a few thermalized configurations, we have measured the values of
$m_0$ where there should be undamped modes in the fifth dimension.
Figure \ref{fig:pbp_vs_ls} shows a measurement of the generalized
chiral condensate as a function of the separation between the source
and sink in the fifth dimension for $m_0$ very close to the
value where a zero of the four-dimensional Wilson operator occurs.

One can clearly see the presence of the expected translationally
invariant mode in the fifth dimension.  However, this mode contributes
about 1\% to the value of the chiral condensate, for this one
lattice where its effect should be largest.  For an ensemble of
lattices, at any given $m_0$ only some will have largely undamped modes.
(This number could be a set of measure zero in the full average,
although this needs to be studied as a function of lattice volume.)
Even if undamped modes do survive the ensemble average, our
example shows that they can make a very small contribution
to an observable.

\section{HADRON MASSES}

To test the formulation in a more practical way, we have measured
hadron masses for quenched domain wall fermions on $8^3 \times 32$
lattices with $L_s = 10, 16, 24$ and 48 for $\beta = 5.7$ and $m_0 =
1.65$.  (Since the conference, we have also simulated at $L_s = 32$.)
We have chosen $m_f$ in the range 0.02 to 0.22.  For $m_f = 0.02$, we
have not gotten good plateaus with 70-90 configurations and
consequently only report results using $m_f > 0.02$.

\begin{figure}[htb]
\epsfxsize=\hsize
\epsfbox{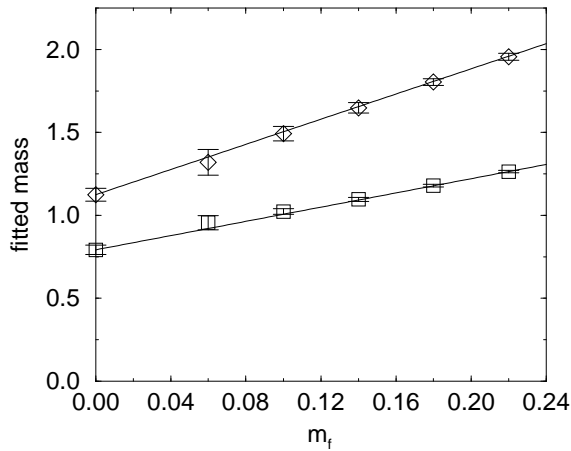}
\caption{Rho ($\Box$) and nucleon ($\Diamond$) masses versus
$m_f$ for 80 quenched $8^3 \times 32$ configurations at
$\beta = 5.7$.  The fifth dimension has $L_s = 48$.}
\label{fig:hdm}
\end{figure}

The rho mass shows no dependence on $L_s$ between 10 and 48, while the
nucleon mass decreases somewhat for larger $L_s$.  For $L_s = 10$,
we find $m_\rho = 0.787(12) + 2.28(5)m_f$ and $m_N = 1.19(3) + 3.71(15)
m_f$.  For $L_s = 48$, we find $m_\rho = 0.792(28) + 2.15(12) m_f$
and $m_N = 1.12(4) + 3.79(19) m_f$.  These are correlated fits to
five or more valence quark masses and all have $\chi^2/dof < 1$.
Figure \ref{fig:hdm} shows the nucleon and rho masses for $L_s = 48$.
The $m_f = 0 $ value for for $m_N/m_\rho$ is 1.42 ($L_s = 48)$,
compared to the staggered fermion values of 
$\approx 1.50$ \cite{bernard}.

We have also measured $m_\pi(m_f)$ for these lattices and find that it
is not a simple linear function of $m_f$ for the full range of quark
masses we used.  Fits including an $m_f^2$ term, or linear fits to the
few smallest values for $m_f$, give values for $m_\pi^2(m_f\rightarrow
0)$ that agree within errors.  We then find that $m_\pi^2(0)$ fits well
to the form $A \exp(-\alpha L_s) + B$, with $B \approx 0.05$.

We have not yet established an explanation for this non-zero intercept
for $m_\pi^2$.  Our studies of the condensate make residual mixing in
the fifth dimension seem unlikely.  Possible causes include the finite
volume effects which give the pion a non-zero intercept in quenched
staggered spectroscopy, a problem with quenching that is brought out by
the zero modes of domain wall fermions or the observed failure of our
integrated $\pi$--$\pi$ correlator to equal $\langle \bar{\psi}
\psi \rangle /m_f$ for small values of $m_f$.

\section{CONCLUSIONS}

Our studies of domain wall fermions for quenched QCD have shown the
technique capable of reproducing expected physics.  The non-zero value
for $m_\pi^2(0)$, even for large $L_s$, deserves more attention.  Since
other formulations have shown similar effects for $m_\pi^2(0)$, we must
check again at weaker coupling and larger volumes before concluding
that this is a new phenomena revealed by domain wall fermions in the
quenched approximation.

The authors would like to thank Robert Edwards, Tim Klassen,
Rajamani Narayanan and Tom Blum for useful discussions.

\end{document}